# The $\phi$ - Meson Physics in the Chiral Model

K.R. Nasriddinov, N.Z. Rajabov and N.E. Iskandarov

Tashkent State Pedagogical University named after Nizami, Yusuf Khos Khojib str.-103, 700100, Tashkent, Uzbekistan

The $\phi \to K^+K^-, K_L^0 K_S^0, K^0 \overline{K}^0$ decays are studied using the method of phenomenological chiral Lagrangians. Calculated values for partial widths for these decay channels are compared with available experimental data.

At present investigations of the $\phi(1020)$ - meson decays in experiments [1] and in the framework of theoretical models [2] are intensively carried out. It should be noted, that such decay channels of the $\phi$ - meson are not considered in chiral models and there are a few attempts [3] to calculate only some radiative decay channels of this meson. Studies of decay channels of this meson in chiral models are of interest because of the following reasons:

First, the $\phi$ - meson mass is comparable to the chiral symmetry breaking scale of 1 GeV, the proposed decays into a rather heavy meson (the $\rho$ or $\omega$ of about 800 MeV) and a light meson (the $\pi$ or $K$) might be manageable along the lines of heavy-mass chiral perturbation theory (similarly to the heavy-baryon chiral perturbation theory of the pion-nucleon sector);

Second, this meson (also another particles) are produced in colliders, for example, at the cooler synchrotron ring (COSY) at the Institut fuer Kernphysic at the Forschungzentrum Juelich in proton-proton collisions. And in this case the environment has a temperature not equal zero. Therefore, we need to take into account the influence of temperature factors of the environment to the decay probabilities.

Well known, that at present we have no any theory for describing weak decay probabilities of hadrons at low energies or large distances. Therefore, there are a lot of phenomenological approaches and models for describing such decay channels of hadrons. One of them is the method of phenomenological chiral Lagrangian's (PCL's) that we used in the past. In the framework of this model we have investigated the weak many-body decays of the $\Lambda_c^+$-baryon, D-mesons and $\tau$ - lepton [4].

In this paper we study the $\phi \to K^+K^-, K_L^0 K_S^0, K^0 \overline{K}^0$ decays by the method of phenomenological chiral Lagrangians (PCL's). Note, that we investigated the isospin - (or G-parity) forbidden decay of the $\phi$ - meson into the $\omega$ - meson and $\pi$ - meson since it is sensitive to the $\phi$-$\omega$-$\rho$- mixing [5]. According to the chiral model this $\phi \to \omega\pi$ decay channel is described, in spite of predictions of another methods [2], by the following diagrams

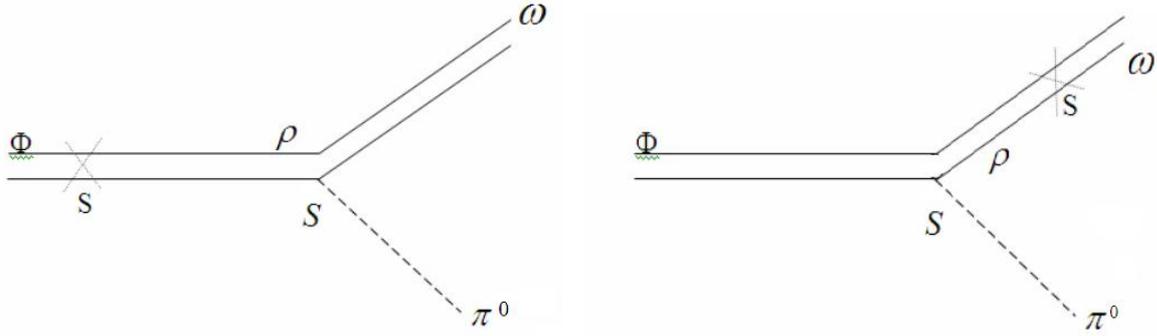

At present the world average [1] for this decay channel is estimated to be

$$Br(\phi \to \omega\pi) = (5.2^{+1.3}_{-1.1}) \times 10^{-5}.$$

In the framework of PCL's the decay rate of this channel with taking into account $\rho(770)$, $\rho(1450)$ and $\rho(1700)$ intermediate meson states have been estimated as

$$\Gamma(\phi \to \omega\pi) = 0.77 \text{ MeV}.$$

For the partial decay rate we have obtained

$$Br(\phi \to \omega\pi) = 0.18,$$

that is four orders of magnitude large than the experiment [1].

In this model the $\phi \to K^+K^-, K^0_L K^0_S, K^0 \overline{K}^0$ decay channels of the $\phi$ - meson are described by the next diagrams

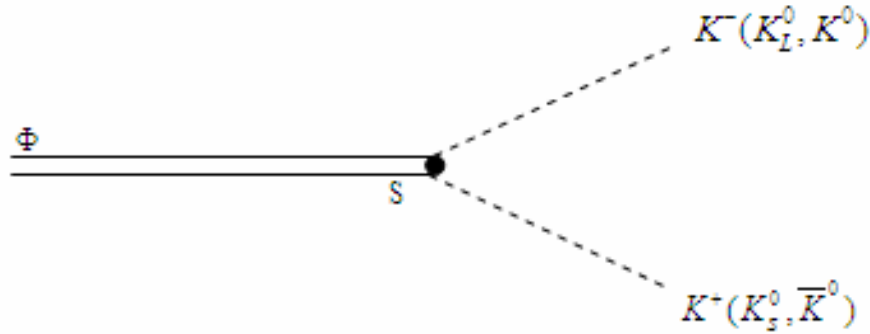

The strong interaction Lagrangian of the $\phi$ meson with $K^-$- and $K^+$- mesons has a form

$$L_S(\phi \to K^-K^+) = -\frac{i}{\sqrt{2}} g \phi_\mu K^- \overset{\leftrightarrow}{\partial} K^+, \quad (1)$$

where, $g$ is the universal coupling constant fixed from the experimental $\rho \to \pi\pi$ decay width $g = \sqrt{12.8\pi}$. The decay amplitude is defined as

$$M = iL_S = i\frac{g}{\sqrt{2}} \varepsilon^\mu(k)(k_{1\mu} - k_{2\mu}), \quad (2)$$

where, $\varepsilon^\mu(k)$ is the polarization vector of the $\phi$ meson; $k_{1\mu}$ - and $k_{2\mu}$ - are the momenta of the $K^+$ - and $K^-$ - mesons, respectively.

The decay probability of the $\phi \to K^+K^-$ channel is defined according to the following expression

$$\widetilde{\Gamma} = \frac{1}{3} \frac{|\overline{M}|^2}{2m_\phi} \Phi, \tag{3}$$

where, $\Phi = \frac{1}{8\pi} \cdot \frac{1}{m_\phi^2} \cdot [m_\phi^2 - 4m_K^2]^{\frac{1}{2}} m_\phi$ is the phase space, $m_\phi$ is the $\phi$ - meson mass.

For the squared decay amplitude we have

$$|\overline{M}|^2 = \frac{g^2}{2}(-k_1^2 + 2k_1 k_2 - k_2^2),$$

where, $k_1^2 = m_K^2$ is the $K^+$ - meson mass; $k_2^2 = m_K^2$ is the $K^-$ - meson mass and

$$k_1 k_2 = \frac{m_\phi - 2m_K^2}{2}.$$

According to (3) the decay rate has the form

$$\Gamma = \frac{12.8}{96 m_\phi^2} [m_\phi^2 - 4m_K^2]^{\frac{3}{2}}.$$

According to this expression we calculated the partial widths for the all $\phi \to K^+ K^-, K_L^0 K_S^0, K^0 \overline{K}^0$ decay channels and have obtained the following values:

$$Br(\phi \to K^- K^+) = 48{,}12\%,$$
$$Br(\phi \to K_L^0 K_S^0) = Br(\phi \to K^0 \overline{K}^0) = 31{,}7\%.$$

Note, that these calculated values for the decay probabilities of these decay channels are in good agreement with experimental data [1]

$$Br^{\exp}(\phi \to K^- K^+) = (49{,}1 \pm 0{,}6)\%,$$
$$Br^{\exp}(\phi \to K_L^0 K_S^0) = (34{,}0 \pm 0{,}5)\%.$$

At present the influence of a density and temperature of a medium to particles properties are intensively studied [6]. Well known that coupling constants of particles are decreased with increasing of a density and temperature of the environment (for example, $g = \sqrt{12{,}8\pi}$ - is the universal coupling constant in the PCL's). In case if we accept that at the $\phi$ - meson level chiral symmetry idea is valid calculated probabilities have a high confidence level. From differences between predicted and experimental decay rates we can conclude about the influence of temperature factors of the environment. In order to take into account this influence we need to have high accuracy experimental data on these decay channels in future.